\documentclass{llncs}
\usepackage{cs_techrpt_cover_crest}

\usepackage[nocompress]{cite}
\usepackage[latin1]{inputenc} 
\usepackage{graphicx}
\usepackage[tight,TABTOPCAP]{subfigure}

\usepackage[pdfpagemode=none,bookmarks=false]{hyperref}

\usepackage{xspace}

\usepackage{ifpdf} 

\makeatletter
\renewcommand{\footnotesize}{\@setfontsize\footnotesize\@viiipt{9.5}}
\makeatother

\newcommand{\smallsec}[1]{\smallskip\noindent{\bf #1.}}

\newcommand{\cpachecker}{{\small\sc CPAchecker}\xspace}
\newcommand{\blast}{{\small\sc Blast}\xspace}
\newcommand{\slam}{{\small\sc Slam}\xspace}

\newcommand{\csisat}{{\small\sc CSIsat}\xspace}
\newcommand{\mathsat}{{\small\sc MathSAT}\xspace}
\newcommand{\simplify}{{\small\sc Simplify}\xspace}

\newcommand{\javabdd}{{\small\sc JavaBDD}\xspace}

\newcommand{\eclipse}{{\small\sc Eclipse}\xspace}

\begin{document}

\title{{\sc CPAchecker}:\\ 
       A Tool for Configurable Software Verification\,%
\thanks{This research was supported by the NSERC grant RGPIN 341819-07.}
}

\subtitle{
(Tool Paper)
}

\author{Dirk Beyer \and M. Erkan Keremoglu}

\reportnumber{SFU-CS-2009-02}
\date{January 30, 2009}

\makecover

\phantom{x}

\author{
Dirk Beyer
\and
M. Erkan Keremoglu
}

\institute{Simon Fraser University, B.C., Canada}

\markboth{Technical Report SFU-CS-2009-02}{Beyer, Keremoglu:
CPAchecker: \\
A Tool for Configurable Software Verification} 

\maketitle

\setcounter{page}{1}
\pagestyle{plain}

\begin{abstract}
Configurable software verification is a recent concept for 
expressing different program analysis and model checking approaches 
in one single formalism.
This paper presents \cpachecker, a tool and framework
that aims at easy integration of new verification components.
Every abstract domain, together with the corresponding operations,
is required to implement the interface of configurable program analysis (CPA). 
The main algorithm is configurable to perform a reachability analysis
on arbitrary combinations of existing CPAs.
The major design goal during the development was to provide a framework for developers
that is flexible and easy to extend.
We hope that researchers find it convenient and productive
to implement new verification ideas and algorithms
using this platform and that it advances the field by
making it easier to perform practical experiments.
The tool is implemented in Java and runs as command-line tool or as \eclipse plug-in.
We evaluate the efficiency of our tool on benchmarks from the software model checker \blast.
The first released version of \cpachecker implements CPAs for predicate abstraction,
octagon, and explicit-value domains.
Binaries and the source code of \cpachecker are publicly available as free software.
\end{abstract}

\section{Overview}

The field of software verification is a fast growing area,
and researchers contribute new ideas and approaches with enormous pace.
The more new approaches are discovered,
the more difficult it is to understand the essential insight
or the fundamental difference that makes a new approach good and better.
Experimental evaluation is often a deciding factor for
whether or not a new approach is considered an advancement of the field.
But it requires a considerable engineering effort to actually build
the software infrastructure for evaluating verification algorithms.
Adapting a suitable parser frontend and transforming the abstract syntax tree into a format
that is convenient for verification algorithms is one example.
The interaction with a theorem prover is yet another issue that
needs to be considered.
There are successful approaches in program analysis
as well as in model checking, but these techniques are rarely combined;
the reason being that it is indeed extremely difficult to combine them.
Most published approaches are not even comparable,
because the choice of the parser frontend, the choice
of the theorem prover, and the choice of the pointer-alias analysis algorithm
in the corresponding tool implementation,
considerably influence the performance and precision of the new verification algorithm.
When evaluating a performance comparison of two approaches,
it is often difficult to identify what the new approach contributes
and what is due to the different environment.
In practice, it was so far extremely difficult to perform
an experimental performance evaluation of one component
while keeping all other components constant.


Configurable program analysis (CPA) provides a conceptual basis
for expressing different approaches in the same formal setting.
The CPA formalism provides an interface for the definition of
program analyses, which includes the abstract domain, the post operator,
the merge operator, and the stop operator~\cite{CPA}.
Consequently, the corresponding tool implementation \cpachecker 
provides an implementation framework
that allows the seamless integration of program analyses that are
expressed in the CPA framework.
The comparison of different approaches 
in the same experimental setting becomes easy and 
the experimental results will be more meaningful (valid).
The tool can be seen as a set of components that are loosely
dependent on each other and that are easy to substitute.


In many respects, \cpachecker is similar to \blast~\cite{BLAST}.
For example, we implemented a predicate abstraction and an explicit-value analysis~\cite{CPAplus}.
However, \blast has several limitations that we need to eliminate,
most prominently, that the architecture and the design are not flexible enough
to implement a pure CPA-based analysis.
As in the \blast project already,
many ideas were taken from \slam~\cite{SLAM}.

The source code, executables, and all benchmark programs
for \cpachecker are available online at
\href{http://www.cs.sfu.ca/~dbeyer/CPAchecker/}
{\small\tt http://www.cs.sfu.ca/$\sim$dbeyer/CPAchecker}.
The tool is free software, released under the Apache 2.0 license.
\cpachecker is an open-source implementation of the
framework of configurable program analysis (CPA).
We hope that other researchers can integrate new techniques for software verification
into \cpachecker and that software-verification technology 
becomes more accessible for practitioners using this platform.


\section{Architecture and Implementation}

\begin{figure}[t]
\centering
\ifpdf
  \includegraphics[scale=0.21]{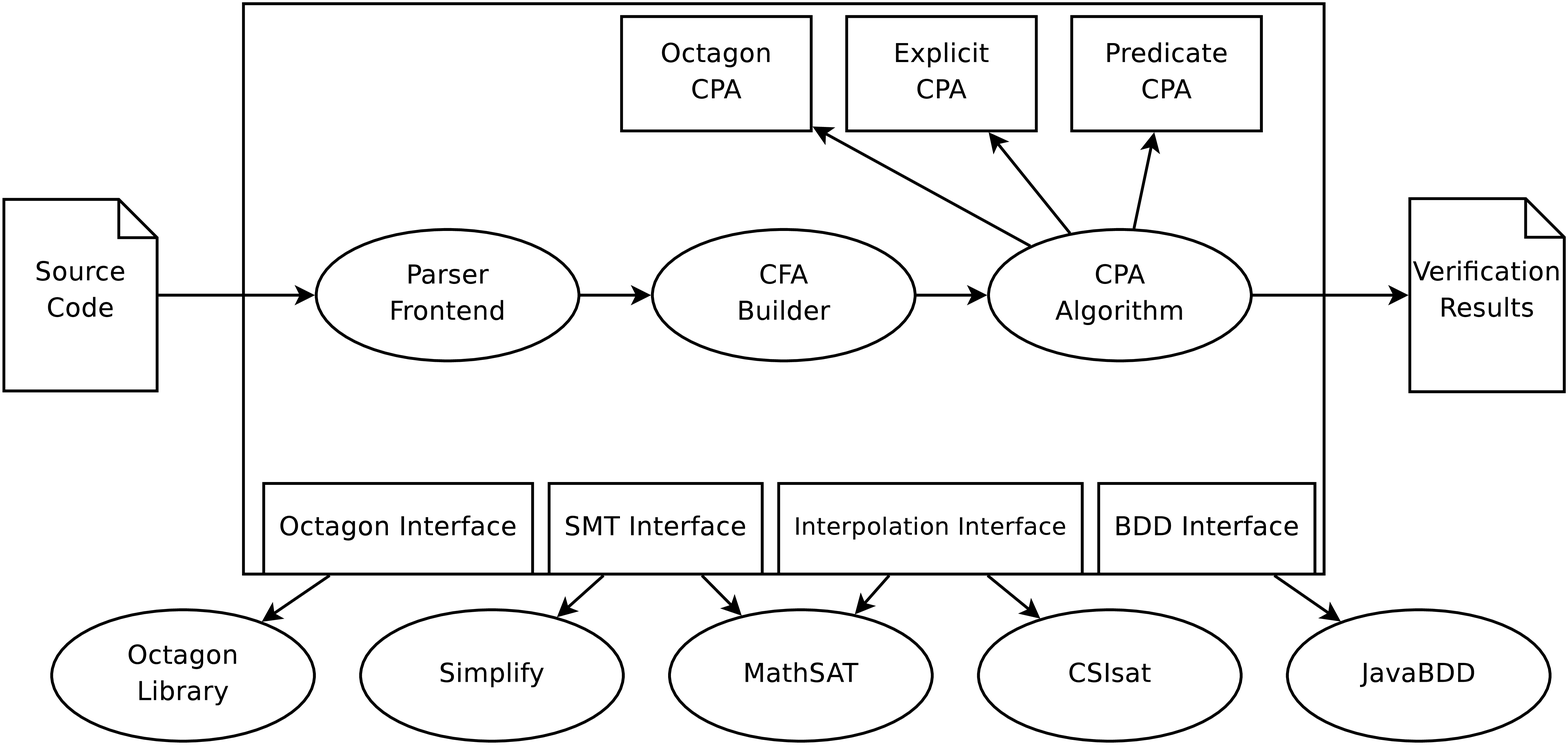}
\else
  \includegraphics[width=12cm]{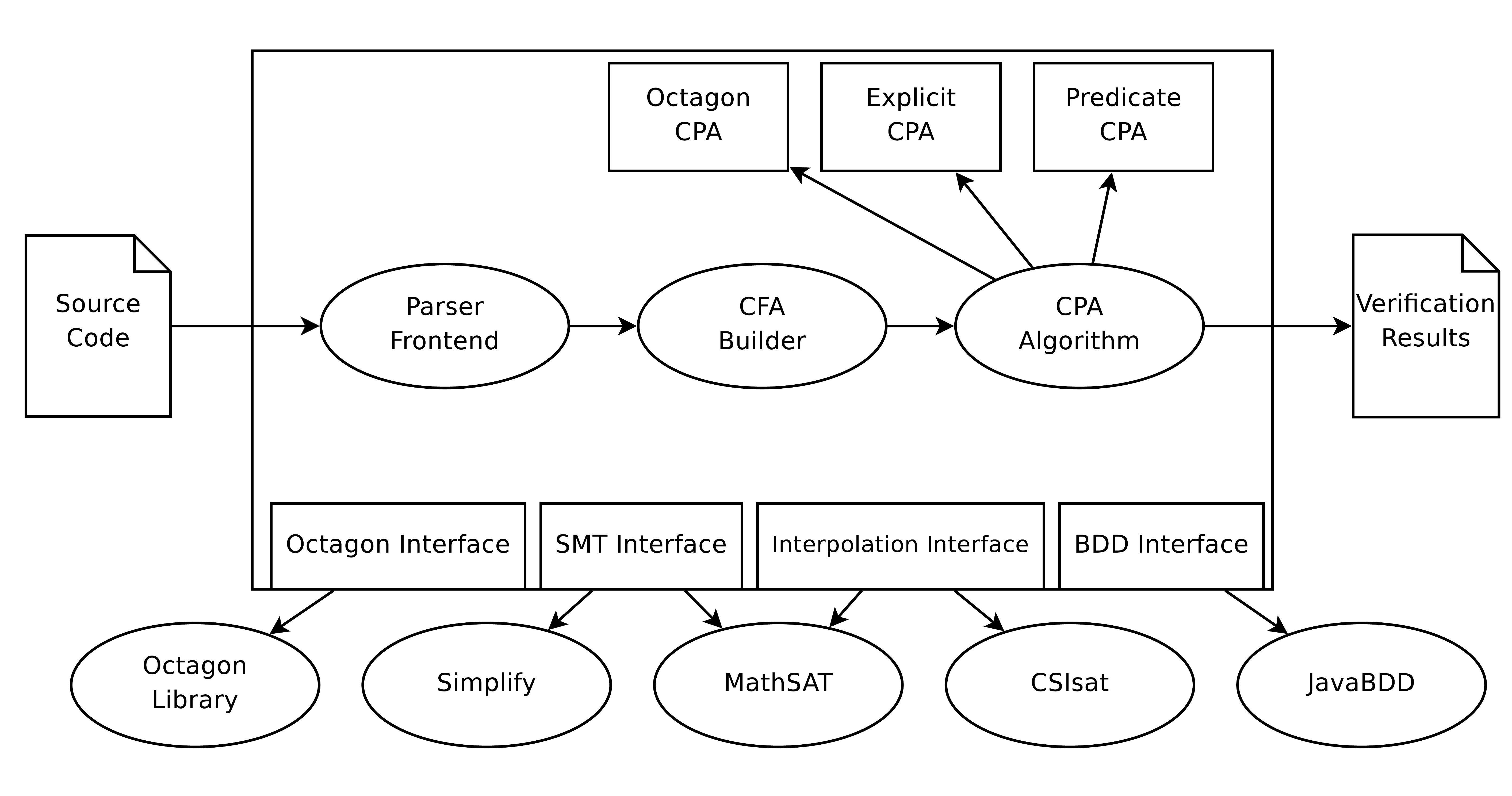}
\fi
\caption{\cpachecker\ --- Architecture overview}
\label{fig:architecture}
\vspace{-2mm}
\end{figure}

Figure~\ref{fig:architecture} shows an overview of the \cpachecker architecture.
The central data structure is a set of control-flow automata (CFA)
(similar to control-flow graphs~\cite{DragonBook}),
which consist of control-flow locations and control-flow edges.
A location represents a program-counter value,
and an edge represents a program operation,
which is either an assume operation, an assignment block,
a function call, or a function return
(we do not consider more complex operations 
 due to a well-known reduction called C intermediate language~\cite{CIL}).
Before a program analysis starts, the input program is 
transformed into a syntax tree, and further into CFAs.
The current version of \cpachecker uses the parser from the CDT 
\footnote{Available at 
\href{http://www.eclipse.org/cdt/}{{\tt http://www.eclipse.org/cdt}}},
a fully functional C and C++ IDE plug-in for the \eclipse platform.
Our framework provides interfaces to SMT solvers and
interpolation procedures, such that the CPA operators 
can be written in a concise and convenient way.
Currently we use \simplify
\footnote{Available at 
\href{http://secure.ucd.ie/products/opensource/Simplify/}
{{\tt http://secure.ucd.ie/products/opensource/Simplify}}}
and \mathsat
\footnote{Available at 
\href{http://mathsat4.disi.unitn.it/}{{\tt http://mathsat4.disi.unitn.it}}}
as SMT solvers,
and \csisat
\footnote{Available at 
\href{http://www.cs.sfu.ca/~dbeyer/CSIsat/}{{\tt http://www.cs.sfu.ca/$\sim$dbeyer/CSIsat}}}
and \mathsat as interpolation procedures.
We use \javabdd
\footnote{Available at 
\href{http://javabdd.sourceforge.net/}{{\tt http://javabdd.sourceforge.net}}}
as BDD package and provide an interface to an Octagon%
\footnote{Available at 
\href{http://www.di.ens.fr/~mine/oct/}{{\tt http://www.di.ens.fr/$\sim$mine/oct}}}
representation as well.

The central algorithm is the program-analysis algorithm
that performs the reachability analysis~\cite{CPA}.
(\cpachecker actually implements CPA+, i.e., CPA with precision adjustment,
 but we skip this detail for better presentation.)
The analysis algorithm operates on an object of the abstract data type CPA,
i.e., the algorithm applies operations from the CPA interface 
without knowing which concrete CPA it is analyzing.
For most configurations, the concrete CPA will be a composite
CPA~\cite{CPA}, which implements the combination of several different CPAs.

\begin{figure}[t]
\centering
\ifpdf
  \includegraphics[scale=0.22]{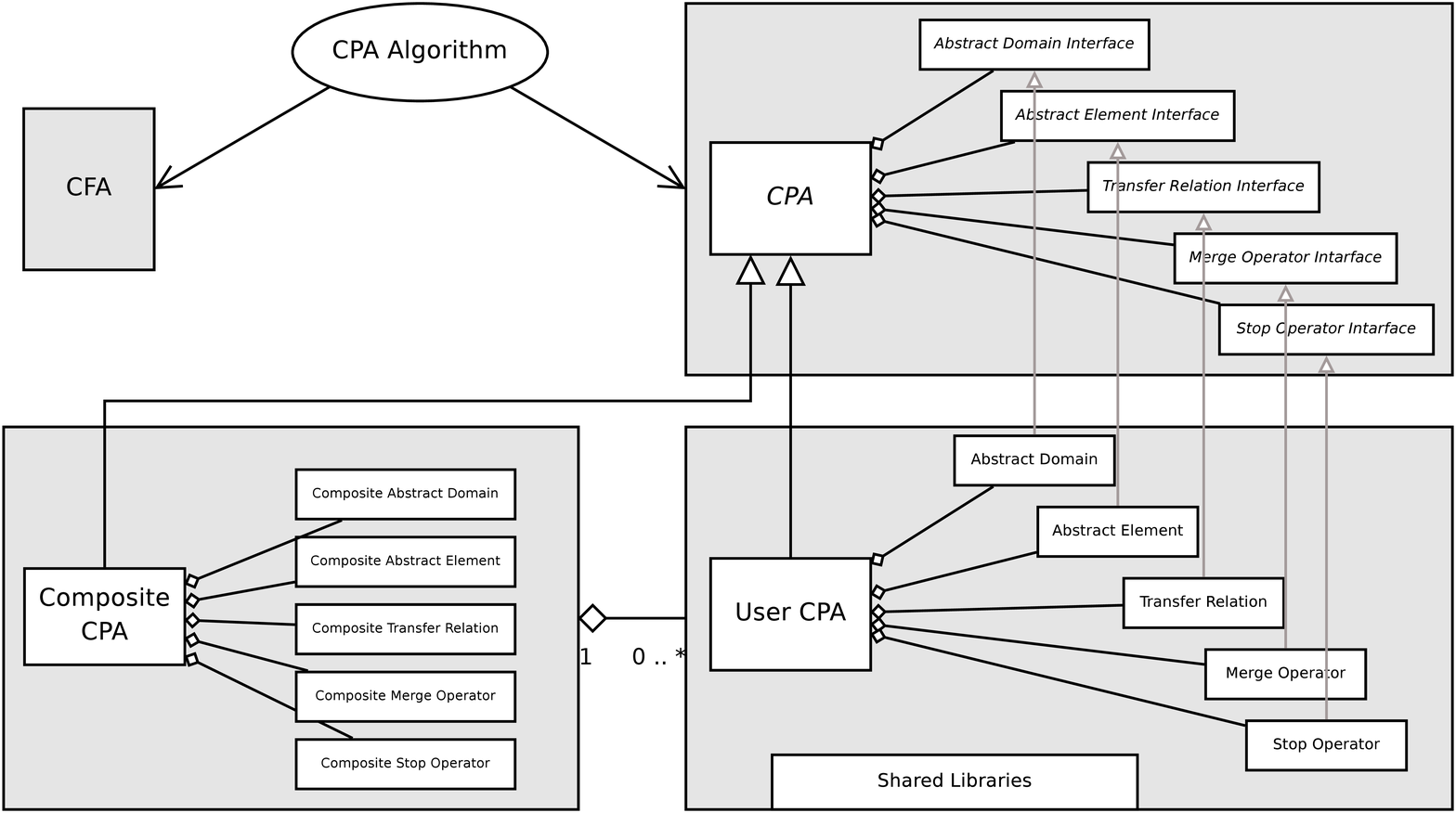}
\else
  \includegraphics[width=12cm]{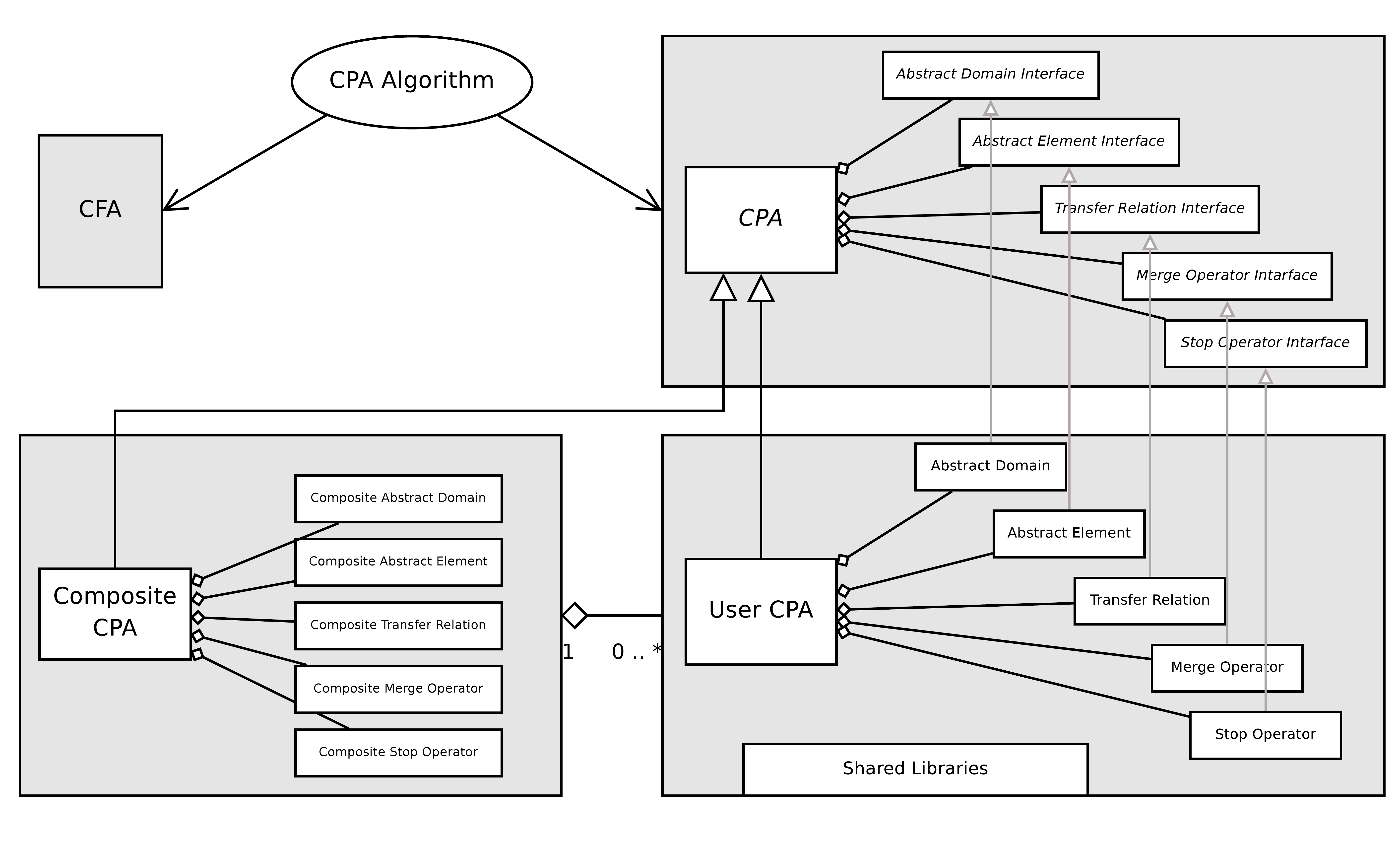}
\fi
\caption{\cpachecker\ --- Design for extension}
\label{fig:design}
\vspace{-2mm}
\end{figure}

In order to extend \cpachecker by integrating an additional CPA
for a new abstract domain, only two steps are necessary.
First, an entry in the global properties file is necessary
in order to announce the new CPA for composition.
Second, the interface for CPA needs to be implemented,
and implementations of all CPA operation interfaces need to be provided.
Figure~\ref{fig:design} shows the interaction:
The CPA algorithm (shown at the top in the figure)
takes as input a set of control-flow automata (CFA) representing the program,
and a CPA, which is in most cases a \emph{Composite CPA}.
The interfaces correspond one-to-one to the formal framework~\cite{CPA}.

The elements in the gray box (top right) in Fig.~\ref{fig:design} 
represent the abstract interfaces of the CPA and the CPA operations.
The two gray boxes at the bottom of the figure
show two implementations of the CPA interfaces,
one is a \emph{Composite CPA} that can combine several other CPAs,
and the other is a \emph{User CPA}.
For example, suppose we want to implement a CPA for shape analysis.
We would provide an implementation for \emph{CPA}, possible called \emph{ShapeCPA},
and implementations for the operation interfaces on the right.
If we want to experiment with several different merge operators,
we would provide several different implementations of
\emph{Merge Operator Interface} that can be freely configured
for use in various experiments.

%
%

\section{Experiments}

\begin{table}
\caption{Performance results; runtime given in seconds of processor time;
         the numbers in the column headings are the threshold values}
\label{tab:experiments}
\centering
\footnotesize
\begin{tabular}{lr@{\hspace{2mm}}r@{\hspace{5mm}}r@{\hspace{5mm}}r@{\hspace{5mm}}rrrrrrrrr}
\hline
Program &                    0 &          2 &              3 &          5 &        $\infty$ \\
\hline

cdaudio\_simpl1 &          $>$1200.00 &       525.90 &      74.65 &      8.43 &      2.96 \\

cdaudio\_simpl1\_BUG ~~&    167.67 &      88.45 &       17.09 &      3.28 &       0.62 \\

diskperf\_simpl1 &         $>$1200.00 &     $>$1200.00 &     36.95 &      21.19 &       280.10 \\

floppy\_simpl3 &           110.38 &      104.02 &      21.94 &      11.91 &       0.88 \\

floppy\_simpl3\_BUG &      42.33 &       37.55 &       7.98 &      2.37 &         0.35 \\

floppy\_simpl4 &           199.22 &      173.92 &      30.17 &      11.22 &       1.43 \\

floppy\_simpl4\_BUG &      42.95 &       36.15 &       8.03 &       2.16 &        0.36 \\

kbfiltr\_simpl1 &          13.77 &        4.59 &       3.50 &        1.02 &       0.42 \\

kbfiltr\_simpl2 &          30.89 &        9.98 &        5.48 &      1.83 &        0.89 \\

kbfiltr\_simpl2\_BUG &     16.17 &        5.76 &        1.24 &      0.73 &        0.32 \\
\hline
\end{tabular}
\vspace{2mm}
\end{table}

\begin{table}
\caption{Statistical data observed during the experiments; 
         a dash indicates that the experiment was aborted after 20\,min;
         'Preds' indicates the number of predicates used in the verification run,
         and 'Refines' indicates the number of refinement steps}
\label{tab:refinement}
\centering
\footnotesize
\begin{tabular}{lr@{\hspace{1mm}}r@{\hspace{3mm}}r@{\hspace{1mm}}r@{\hspace{3mm}}r@{\hspace{1mm}}r@{\hspace{3mm}}r@{\hspace{1mm}}r}
\hline
Program&                  \multicolumn{2}{c}{0} &   \multicolumn{2}{c}{2} &   \multicolumn{2}{c}{3} &   \multicolumn{2}{c}{5}\\
&                         Preds &    Refines &      Preds &    Refines &      Preds &    Refines &      Preds &    Refines \\
\hline
cdaudio\_simpl1 &            - &          - &         81 &        332 &         12 &         76 &          2 &         11  \\

cdaudio\_simpl1\_BUG ~~&   112 &        242 &         56 &        140 &         12 &         38 &          2 &         10  \\

diskperf\_simpl1 &           - &          - &          - &          - &         20 &         61 &          4 &         34 \\

floppy\_simpl3 &            81 &        219 &         51 &        167 &         20 &         51 &          4 &         21 \\

floppy\_simpl3\_BUG &       47 &        125 &         38 &         93 &         13 &         28 &          6 &          5 \\

floppy\_simpl4 &            96 &        307 &         54 &        219 &         20 &         58 &          4 &         19 \\

floppy\_simpl4\_BUG &       47 &        125 &         38 &         93 &         13 &         28 &          6 &          5 \\

kbfiltr\_simpl1 &           30 &         70 &          7 &         22 &          5 &         11 &          1 &          2 \\

kbfiltr\_simpl2 &           48 &        133 &          7 &         40 &          5 &         11 &          1 &          2 \\

kbfiltr\_simpl2\_BUG &      44 &         89 &         16 &         34 &          1 &          4 &          0 &          1 \\
\hline
\end{tabular}
\vspace{-1mm}
\end{table}

We report experiments in order to demonstrate that the tool implementation
performs reasonable well on well-known benchmark examples.
We pick a configuration for program analysis that was previously used~\cite{CPAplus},
namely, the combination of an explicit-value analysis and a predicate-abstraction.
Explicit-value analysis, also known as constant propagation, keeps
track of values of integer variables.
The predicate abstraction is based on Cartesian abstraction
and lazy abstraction~\cite{LazyAbstraction}.
We run the analysis on various verification problems 
for simplified versions of Windows device drivers.
The verification property is always a safety property
(reachability of a certain error location under certain variable values)
and is thus contained in the source code.
The same program name ending with a different number
indicates that the same program is present with a different
simplification applied to the source code.
If the program name ends with ``BUG'' then
a defect was artificially introduced into the program.

The overall performance results obtained in our initial development phase 
of \cpachecker are satisfactory, although
optimization was not the main design goal --- rather we focussed on
a portable and flexible environment to be used for many different analysis purposes.
All experiments were performed on a GNU/Linux (Ubuntu 8.10) 
x86\_32 machine with an Intel Core 2 Duo processor and 2\,GB RAM.
We limited the memory for the Java virtual machine to 1.8\,GB and
set the time limit for termination to 1200\,s.


Table~\ref{tab:experiments} shows the performance results for different configurations.
The first column of the table lists the names of the programs.
The next five columns report the runtimes for the analysis configuration where predicate abstraction
and explicit-value analysis are used together.
The threshold (the number in the column heading) 
indicates how many different explicit values where tracked for each variable 
(cf.~\cite{CPAplus} for the details).
After reaching this threshold the value of the variable is set to~$\top$, i.e.,
nothing can be said about the value of the variable in the explicit analysis.
This might lead to an infeasible path and the predicate-abstraction domain discovers predicates
in order to track the missing variables and to eliminate the infeasible program path.
We experimented with five different threshold values,
where $0$ represents the extreme case of pure predicate abstraction-based analysis,
and $\infty$ represents the extreme case of pure explicit-value analysis.
Table~\ref{tab:experiments} indicates that the best performance
in total for this set of programs is achieved with a threshold of~$5$,
which represents a good tradeoff between the expensive but abstract predicate abstraction
and the simple but exploding explicit-value analysis.
It is interesting to observe that pure predicate abstraction is not tractable
for some of the experiments (time out reached).

Table~\ref{tab:refinement} shows the number of predicates and the number of refinement iterations
needed to obtain the verification result.
Surprisingly, many facts can be tracked by explicit values,
and thus the number of predicates in the abstract-successor computations is drastically reduced.
Also, the number of refinements that are necessary to discover predicates
is significantly reduced
(note that many different refinements might discover the same predicate for different locations).

%


\bigskip
\smallsec{Acknowledgments}
We thank Tom Henzinger, Ranjit Jhala, and Rupak Majumdar for the
fruitful collaboration in the \blast project.
\blast served as example for \cpachecker in several aspects.
We also thank Alberto Griggio, Andreas Holzer, and Michael Tautschnig
for their valuable comments
and for their code contributions to \cpachecker.
We thank Alberto especially for his contribution to the predicate-abstraction analysis.


\end{document}